\documentclass[12pt]{article}

\usepackage{sbc-template}

\usepackage{graphicx,url}

\usepackage[brazil]{babel}   
\usepackage{booktabs}
\usepackage{amsmath}
\usepackage{listings}
\usepackage{xcolor}
\usepackage{float}
\usepackage{multicol}
\usepackage{tabularx}
\usepackage{array}

\def\instnum{}

\title{Semantic Clustering of Civic Proposals: A Case Study on Brazil's National Participation Platform}

\author{
Ronivaldo Ferreira\inst{1}, Guilherme da Silva\inst{2}, \\ Carla Rocha\inst{2}, Gustavo Pinto\inst{1}%
}

\address{
Faculdade da Computação\\ Universidade Federal do Pará (UFPA)\\
 Belém -- PA -- Brazil
\nextinstitute
 Faculdade do Gama \\
 Universidade de Brasília (UnB)\\
 Brasília -- DF -- Brazil
\email{ronivaldo.junior@icen.ufpa.br}
}

\begin{document} 

\maketitle

\begin{abstract}
Promoting participation on digital platforms such as Brasil Participativo has emerged as a top priority for governments worldwide. However, due to the sheer volume of contributions, much of this engagement goes underutilized, as organizing it presents significant challenges: (1) manual classification is unfeasible at scale; (2) expert involvement is required; and (3) alignment with official taxonomies is necessary. In this paper, we introduce an approach that combines BERTopic with seed words and automatic validation by large language models. Initial results indicate that the generated topics are coherent and institutionally aligned, with minimal human effort. This methodology enables governments to transform large volumes of citizen input into actionable data for public policy.
\end{abstract}

\begin{resumo}
Promover a participação em plataformas digitais, como o Brasil Participativo, tornou-se uma prioridade para governos em todo o mundo. No entanto, devido ao enorme volume de contribuições, grande parte desse engajamento fica subutilizada, pois sua organização apresenta desafios significativos: (1) a classificação manual é inviável em escala; (2) é necessária a participação de especialistas; e (3) é preciso o alinhamento com taxonomias oficiais. Neste artigo, apresentamos uma abordagem que combina BERTopic com palavras-semente e validação automática realizada por grandes modelos de linguagem. Resultados iniciais indicam que os tópicos gerados são coerentes e alinhados institucionalmente, com esforço humano mínimo. Essa metodologia permite aos governos transformar grandes volumes de contribuições dos cidadãos em dados acionáveis para a formulação de políticas públicas.
\end{resumo}

\section{Introdução}

A promoção da participação digital emergiu como uma agenda prioritária para governos em escala global, refletindo a crescente necessidade de modernização dos processos democráticos. Nesse contexto, surgem iniciativas voltadas à criação de plataformas tecnológicas que viabilizam a coleta, organização e análise de propostas da sociedade civil para a formulação de políticas públicas. Essas plataformas buscam não apenas ampliar o acesso dos cidadãos aos processos decisórios, mas também fortalecer a transparência, a legitimidade e a responsividade das ações governamentais.

Nesse contexto, surgiu em 2023 a plataforma Brasil Participativo (BP)~\cite{aguiar2024colaboraccao}, com foco na ampliação do debate público mediante participação digital. A adoção dessa plataforma visou operacionalizar a iniciativa Plano Plurianual Participativo (PPA), resultando em uma participação massiva. Posteriormente, foram executados processos participativos de grande relevância política e social. O conjunto desses processos gerou mais de 1,4 milhão de cidadãos cadastrados e mais de 8 mil propostas elaboradas pelos cidadãos participantes (mais na Seção~\ref{sec:bp}).


No processo de formulação de políticas públicas, a classificação das propostas apresentadas pela sociedade civil é crucial para definir tanto as ações a serem implementadas quanto os responsáveis por sua execução~\cite{clemente2018leonardo, saravia2007politicas}. Todavia, esse procedimento manual em cenários de participação digital de larga escala enfrenta duas limitações que comprometem sua eficácia. Primeiramente, o volume expressivo de contribuições, que pode atingir dezenas ou centenas de milhares de propostas. Além disso, a classificação depende de conhecimentos especializados e multidisciplinares, abrangendo desde as especificidades temáticas de cada área de governo até os marcos regulatórios e as estruturas organizacionais, o que eleva custos temporais e financeiros.

Nesse contexto, técnicas de processamento de linguagem natural podem viabilizar o processamento em escala do acervo do BP, garantindo agilidade, consistência e rastreabilidade na categorização. Essas soluções apoiam o mapeamento semântico em taxonomias institucionais e favorecem a incorporação estruturada dos resultados nos ciclos de elaboração de políticas públicas.


Desenvolvemos um pipeline de extração de tópicos baseado em BERTopic~\cite{grootendorst2022bertopic} para organizar e interpretar o corpus. Além da abordagem não supervisionada, investigamos duas estratégias semi-supervisionadas: (i) \emph{palavras-semente}, que incorporam termos extraídos do corpus para orientar a formação de clusters segundo categorias institucionais; e (ii) tópicos guiados, que impõem rótulos predefinidos para construir uma hierarquia semântica. Um Modelo de Linguagem de Grande Escala (LLM) foi empregado para validar automaticamente os tópicos e gerar interpretações, reduzindo a intervenção manual. A qualidade dos modelos foi medida por métricas de coerência, diversidade e alinhamento com a taxonomia.

Este estudo é guiado pelas seguintes questões de pesquisa:

\begin{enumerate}
\item \textbf{RQ1.} Quais ajustes nos parâmetros do BERTopic maximizam a coerência semântica e a diversidade temática dos tópicos extraídos?
\item \textbf{RQ2.} Em que grau a incorporação de \emph{palavras-semente} do VCGE fortalece o alinhamento semântico com as categorias oficiais?
\end{enumerate}

Este artigo apresenta a plataforma Brasil Participativo e seu vocabulário (Seção 2), a metodologia de modelagem de tópicos e validação (Seção 3), os resultados e respostas às questões de pesquisa (Seção 4), discussão sobre impactos e evolução (Seção 5), trabalhos relacionados (Seção 6) e conclusão com direções futuras (Seção 7). Artefatos disponíveis em: \url{https://github.com/BERTopic/bertopic_bp}.

\section{A Plataforma Brasil Participativo}\label{sec:bp}

A plataforma Brasil Participativo (BP) foi instituída em 2023 pela Secretaria-Geral da Presidência da República como ambiente digital integrado para coleta, organização e priorização de contribuições cidadãs ao Plano Plurianual (PPA) 2024–2027, instrumento previsto na Constituição Federal\footnote{\url{https://www.planalto.gov.br/ccivil_03/_ato2023-2026/2024/lei/L14802.htm}}. Durante a execução do processo participativo, a plataforma registrou cerca de 1,4 milhão de acessos e mais de 8.000 propostas submetidas pelos cidadãos\footnote{\url{https://www.gov.br/planejamento/documentos-hospedados-para-gerar-qrcodes/relatorio-ppaparticipativo}}.


A plataforma apoia-se nas experiências federais de participação social, organizando-se em três frentes institucionais (planos, conferências e consultas) e oferecendo, para cada uma, ferramentas de propostas, enquetes e eventos. Dentre elas, as propostas despontaram como o principal canal de engajamento. O Brasil Participativo unifica esses instrumentos num portal colaborativo e interinstitucional, reunindo órgãos governamentais, Dataprev, UnB e a comunidade Decidim-Brasil.

\subsection{Vocabulário Controlado de Governo Eletrônico (VCGE)}
\label{sec:Context-VCGE}


O VCGE é a taxonomia oficial mantida pela Secretaria de Logística e Tecnologia da Informação do Ministério do Planejamento, Orçamento e Gestão, criada para uniformizar a indexação de conteúdos informacionais no âmbito do Governo Federal\footnote{\url{https://www.gov.br/governodigital/pt-br/infraestrutura-nacional-de-dados/registros-de-referencia/vocabulario-controlado-do-governo-eletronico}}. Está organizado em um nível superior (N1) com 26 domínios temáticos, cada um subdividido em termos de segundo nível (N2); a Tabela~\ref{tab:vcge-categorias} apresenta dez desses domínios como exemplo, e a lista completa está disponível no repositório do projeto. Cada conceito recebe um identificador numérico único e uma URI estável, assegurando interoperabilidade semântica e facilitando a integração de bases de dados, portais de consulta pública e APIs governamentais. Em plataformas de participação digital como a BP, o VCGE mapeia as propostas dos cidadãos em linguagem institucional consensual, reduzindo ambiguidades e inconsistências, permitindo referências precisas em relatórios, portais de transparência e sistemas de gestão de políticas, e apoiando a agregação e análise de dados.

\begin{table}[ht]
  \centering
  \scriptsize
  \caption{Mapeamento resumido de 10 categorias do VCGE (versão 2.1.0)}
  \label{tab:vcge-categorias}
  \begin{tabular}{@{}p{0.25\textwidth} p{0.63\textwidth}@{}}
    \toprule
    \textbf{Nível 1 (N1)} & \textbf{Nível 2 (N2)} \\
    \midrule
    Agropecuária e Pesca & Defesa e vigilância sanitária, Produção agropecuária, Outros em Agropecuária \\
    Comércio e Serviços          & Comércio externo, Defesa do Consumidor, Turismo, Outros em Comércio e Serviços \\
    Comunicações                 & Comunicações Postais, Telecomunicações, Outros em Comunicações \\
    Cultura                      & Difusão Cultural, Patrimônio Cultural, Outros em Cultura \\
    Defesa Nacional              & Defesa Civil, Defesa Militar, Outros em Defesa Nacional \\
    Energia                      & Combustíveis, Energia elétrica, Outros em Energia \\
    Esporte e Lazer              & Esporte comunitário, Esporte profissional, Lazer, Outros em Esporte e Lazer \\
    Habitação                    & Habitação Rural, Habitação Urbana, Outros em Habitação \\
    Indústria                    & Mineração, Produção Industrial, Propriedade Industrial, Outros em Indústria \\
    Saneamento                   & Saneamento Básico Rural, Saneamento Básico Urbano, Outros em Saneamento \\
    \bottomrule
  \end{tabular}
\end{table}

\section{Metodologia}

\begin{figure}[ht]
  \centering
  \includegraphics[width=\linewidth]{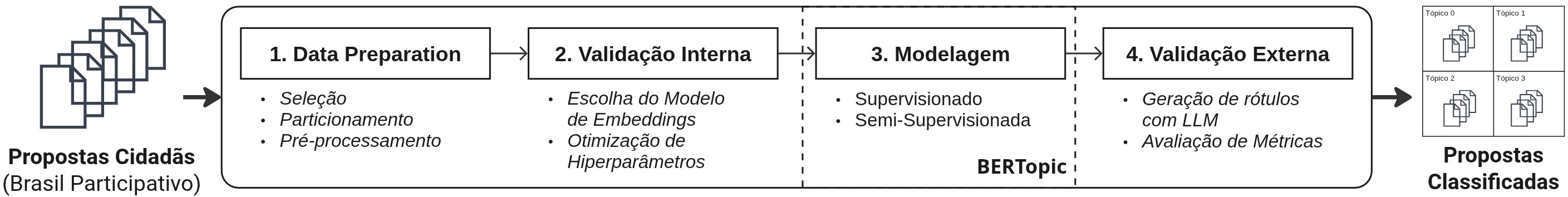}
  \caption{Pipeline de categorização temática com BERTopic}
  \label{fig:metodo}
\end{figure}

O desenho metodológico (Fig.~\ref{fig:metodo}) engloba quatro etapas: (i) extração e pré-processamento de dados; (ii) validação interna e ajuste de hiperparâmetros; (iii) modelagem de tópicos, mesclando descoberta não supervisionada e refinamento semi-supervisionado por tópicos-semente; e (iv) validação externa, comparando clusters ao VCGE via métricas de concordância e análise qualitativa do LLM.

\subsection{Seleção, Particionamento e Pré-processamento de Dados}\label{sec:metodologia-dados-preprocessamento}
Foram extraídas 10.186 propostas dos processos “Plano Clima”, “Plano Plurianual Participativo” e “Congresso da Juventude”. Após remoção de 164 duplicatas e registros vazios, restaram 10.022 amostras, divididas em 80\% para treino (8.014) e 20\% para teste (2.008). Para preservar a representação temática entre os conjuntos, o primeiro autor realizou uma análise manual das categorias selecionadas pelos proponentes na plataforma.

No conjunto de treino, o LLM Gemma 3 (12 bilhões de parâmetros)\footnote{\url{https://ai.google.dev/gemma/docs/core}} gerou rótulos automáticos nos níveis N1 (categorias gerais) e N2 (subcategorias), adotados como \emph{gold standard} na etapa de avaliação. O pré-processamento, aplicado a todo o corpus, incluiu normalização para minúsculas, remoção de pontuação e espaços duplicados, eliminação de \emph{stopwords} em português, tokenização e lematização (por exemplo, \texttt{cidadãos} reduzido a \texttt{cidadão}) para evitar que variações morfológicas fossem tratadas como termos distintos.

\subsection{Validação Interna}
\label{sec:metodologia-validacao-interna}

Nesta etapa, buscamos garantir que tanto a representação semântica quanto a configuração de tópicos estejam ajustadas ao nosso corpus. Seguimos o baseline de validação conforme~\cite{hott2023evaluating}, dividindo o trabalho em duas fases: (i) escolha do modelo de embeddings mais adequado e (ii) otimização dos hiperparâmetros do BERTopic.

\paragraph{Escolha do Modelo de Embeddings:}
Inicialmente, avaliamos quatro modelos de embeddings (ver Tabela~\ref{tab:modelos_embeddings}). O conjunto de treino foi submetido a cada um desses modelos, gerando vetores de alta dimensão que capturam a semântica de cada proposta. Em seguida, para cada conjunto de embeddings, instanciamos e executamos o BERTopic variando o número de tópicos-alvo (\texttt{nr\_topics}) nos valores 10, 30, 50, 70, 90, 110 e 130, além do modo \texttt{auto}, que determina automaticamente o melhor número de temas. Cada combinação foi repetida dez vezes para avaliar a consistência dos resultados. Em cada execução, calculamos as métricas Coerência Normalizada (NC) e Diversidade Normalizada (ND). Com esses dois indicadores, definimos a Pontuação Ponderada (WS), como: \[
    WS \;=\; 0{,}8 \times NC \;+\; 0{,}2 \times ND.
\]

\begin{table}[htbp]
  \centering
  \caption{Modelos de Embeddings Avaliados}
  \label{tab:modelos_embeddings}
  \resizebox{0.9\linewidth}{!}{
  \begin{tabular}{@{} p{5cm} p{0.65\textwidth} @{} }
    \toprule
    \textbf{Modelo} & \textbf{Descrição} \\
    \midrule
    LaBSE \cite{feng2020language} &
      Modelo multilíngue BERT (109 idiomas), com treino agnóstico ao idioma para representar semântica com alta qualidade. \\[6pt]
    Sentence-BERT \cite{reimers2019sentence} &
      Duas variantes do Sentence-BERT para múltiplos idiomas (MiniLM-L12-v2 e mpnet-base-v2), ambas otimizadas para similaridade semântica. \\[6pt]
    BERTimbau \cite{souza2020bertimbau} &
      Modelos PT-BR base e large, treinados em corpus nativo; large foca em capacidade, base em eficiência. \\[6pt]
    LegalBERTPT-br \cite{silva2021evaluating} &
      Modelo jurídico em PT-BR, com SimCSE sobre BERTimbau, focado em nuances semânticas legais. \\
    \bottomrule
  \end{tabular}
  }
\end{table}

\paragraph{Otimização de Hiperparâmetros.}
Após definir o modelo de embeddings, otimizamos os principais hiperparâmetros do BERTopic por meio de busca em grade. Testamos dois intervalos de $n$-gramas (unigramas e bigramas) além de variações em \texttt{nr\_topics} (entre 10 e 130, incluindo o modo \texttt{auto}). Paralelamente, investigamos diferentes limites para o parâmetro \texttt{min\_topic\_size}, estabelecendo os valores 3, 5, 10, 15, 20, 25 documentos por tópico, a fim de evitar tópicos com poucas amostras ou temas genéricos demais.


\paragraph{Personalização de Parâmetros ao Contexto do Corpus.}
Em~\cite{hott2023evaluating}, os intervalos de \texttt{nr\_topics} foram (10, 13, 14, 15, 16, 17, 19, 20, auto) e os de \texttt{min\_topic\_size}, de 10 a 100 em passos de 10. Optamos por faixas distintas por duas razões: (1) o baseline usou documentos longos (licitações em PDF), enquanto trabalhamos com textos curtos, muitas vezes parágrafos; e (2) nossa plataforma abrange diversos processos, como PPA, G20 e Plano Clima, além de 55 categorias selecionáveis. Essa combinação de concisão e diversidade temática tende a reduzir significativamente o número de tópicos extraídos quando se utilizam valores elevados de \texttt{min\_topic\_size} (\textgreater 30), comprometendo a granularidade e a representatividade dos temas.

\subsection{Modelagem de Tópicos}
\label{sec:metodologia-topicmodel}
Utilizamos uma abordagem não supervisionada e uma estratégia semi-supervisionada, que incorpora conhecimento institucional extraído do VCGE.

\subsubsection{Abordagem Não Supervisionada}
\label{sec:metodologia-naosupervisionado}
No modelo não supervisionado, iniciamos com o conjunto de documentos pré-processados e os embeddings BERTimbau-large correspondentes. Ao instanciar o BERTopic, definimos explicitamente \(\texttt{min\_topic\_size} = 10\), \(\texttt{nr\_topics} = 70\) e \(\texttt{n\_gram\_range} = (1,1)\). O método \texttt{fit\_transform} foi então aplicado, resultando em dois vetores principais: \texttt{topics\_train}, que contêm o rótulo de tópico atribuído a cada documento, e \texttt{probs\_train}, que indica a confiança do modelo em cada atribuição. Após a inferência, construímos uma tabela detalhada que relaciona cada tópico ao número de documentos atribuídos, ao nome (gerado automaticamente a partir das palavras mais representativas) e à lista de palavras-chave que caracterizam aquele tema. Essa tabela serve de base para análises posteriores, permitindo identificar quais assuntos emergiram sem qualquer orientação prévia.

\subsubsection{Abordagem Semi-Supervisionada}
\label{sec:metodologia-semisupervisionado}

Este processo foi conduzido em três etapas.  
Na primeira etapa, construímos um dicionário \texttt{VCGE\_TAXONOMY}, em que cada chave corresponde a uma categoria N1 e cada valor é a lista de suas subcategorias N2 (ver Tabela~\ref{tab:vcge-categorias}). Na segunda etapa, para cada categoria, geramos uma lista reduzida de \texttt{seed\_words}, composta pelo próprio nome da categoria e por até cinco de suas subcategorias. Também removemos sistematicamente os termos \texttt{Outros}, pois não trazem contribuição semântica relevante para o alinhamento.

Os termos N2 foram escolhidos para representar de forma concisa cada domínio de N1, garantindo que as palavras mais representativas recebam peso diferenciado no cálculo de relevância do nosso \emph{c-TF–IDF} personalizado. Por exemplo, a categoria \texttt{administração} foi resumida em \texttt{compras governamentais}, \texttt{orçamento}, \texttt{patrimônio}, \texttt{serviços públicos} e \texttt{recursos humanos}, enquanto \texttt{economia e finanças} ficou exemplificada por \texttt{defesa da concorrência}, \texttt{política econômica} e \texttt{sistema financeiro}, e assim sucessivamente para as demais categorias.  

Por fim, na terceira etapa, instanciamos um objeto \texttt{ClassTfidfTransformer} personalizado, definindo \(\texttt{seed\_multiplier} = 2\) para reforçar o peso das \texttt{seed\_words} no cálculo de relevância. Por fim, configuramos o BERTopic com os parâmetros \(\texttt{min\_topic\_size} = 10\), \(\texttt{nr\_topics} = 70\) e \(\texttt{n\_gram\_range} = (1,1)\), além de incluir as variáveis \texttt{ctfidf\_model} e \texttt{seed\_topic\_list}, que contêm, respectivamente, o transformer personalizado e o conjunto completo de tópicos-semente.

\subsection{Validação Externa}
\label{sec:metodologia-validacao}

A validação externa do pipeline de análise textual concentrou-se em duas avaliações complementares. A primeira consistiu na rotulação automática das propostas pelo LLM Gemma 3:12B, enquanto a segunda mediu a aderência dos tópicos inferidos pelo BERTopic em relação aos rótulos oficiais do VCGE.

\subsubsection{Rotulação Automática com LLM}
\label{sec:metodologia-llm}

Para automatizar a atribuição de rótulos, carregamos duas listas com os termos válidos do VCGE (\texttt{vcge\_n1\_option} e \texttt{vcge\_n2\_option}). Em seguida, construímos um prompt com a sequência das opções de nível 1, de nível 2 e até 1.500 caracteres do documento, truncados para preservar a coerência. O prompt instruía o modelo Gemma 3:12B a responder com dois termos exatos, um de cada nível, no formato \texttt{<nível 1>, <nível 2>}, sem informações adicionais.

\begin{quote}
\ttfamily
CLASSIFIQUE este texto usando APENAS UM destes termos oficiais do VCGE: 
\$termos\_VCGE.

TEXTO: \$text\_limited\_in\_1500\_chars
  
RESPONDA APENAS com o termo exato. Nada além.
\normalfont
\end{quote}

Utilizamos \texttt{temperature=0.2} e \texttt{num\_ctx=2048}. Quando o modelo sugeria termos fora das listas, o rótulo \texttt{no\_match} era atribuído. As respostas foram adotadas como \emph{gold standard} para comparação com métodos não supervisionados e semi-supervisionados. Para facilitar a interpretação dos tópicos, empregamos o GPT o3-mini em prompt zero-shot para nomear os clusters extraídos pelo BERTopic. O modelo foi escolhido por sua capacidade de reasoning. O tópico "-1" era rotulado como "Outliers"; os demais, com rótulos claros de até três palavras. A saída esperada era um dicionário com os IDs dos tópicos como chaves e seus respectivos rótulos como valores.




\subsubsection{Avaliação de Tópicos}
\label{sec:metodologia-avaliacao-topicos}

Para avaliar a qualidade dos tópicos gerados pelo BERTopic, utilizamos o conjunto de teste (20\% dos documentos) com o mesmo pré-processamento descrito na Seção~\ref{sec:metodologia-dados-preprocessamento}. Em seguida, aplicamos os modelos não supervisionado e semi-supervisionado, obtendo \texttt{topics\_test} e \texttt{probs\_test}. Outliers (\texttt{topic} = –1) foram filtrados, e para cada proposta registramos: texto limpo, rótulos do LLM (\texttt{VCGE\_N1} e \texttt{VCGE\_N2}), tópicos inferidos e suas probabilidades.

Para quantificar a concordância com os rótulos oficiais, calculamos as métricas Adjusted Rand Index (ARI) e Normalized Mutual Information (NMI), separadamente para os níveis N1 e N2. Adicionalmente, construímos matrizes de contingência entre \texttt{topic} e rótulo de referência, complementadas por heatmaps com contagens absolutas e proporções normalizadas por categoria. Essas visualizações indicaram o grau de alinhamento entre os tópicos e as categorias do VCGE, além de possíveis lacunas ou sobreposições temáticas, oferecendo subsídios para aprimoramentos no pipeline de classificação.

\section{Resultados}

\subsection{RQ1 – Quais ajustes nos parâmetros do BERTopic maximizam a coerência semântica e a diversidade temática dos tópicos extraídos?}

\begin{figure}[ht]
  \centering
  \includegraphics[width=0.75\linewidth]{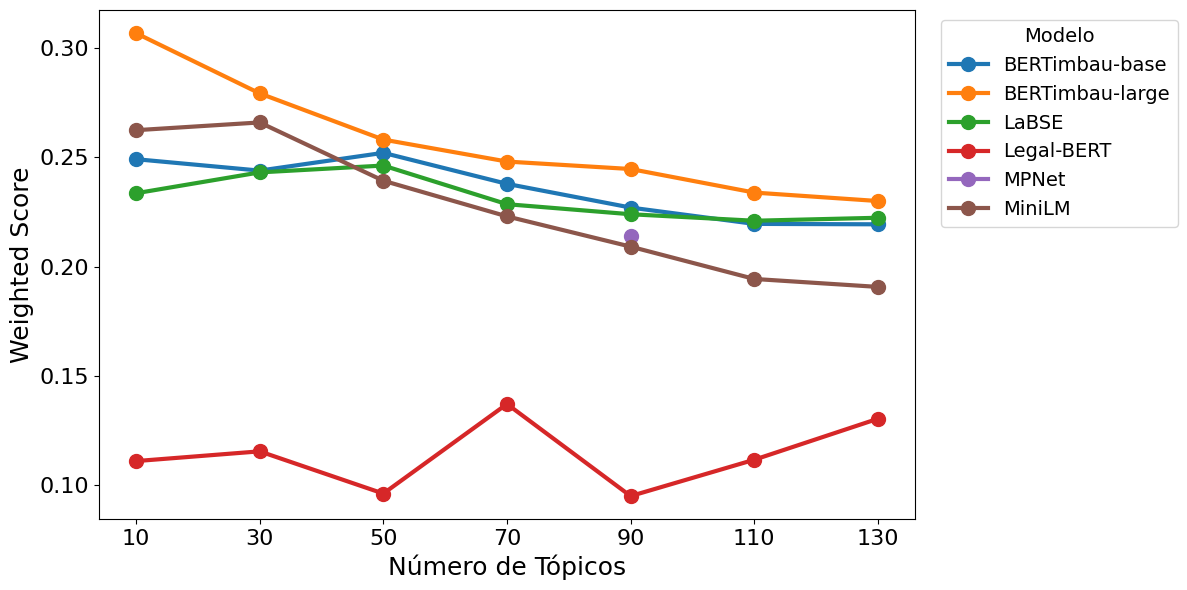}
  \caption{Weighted Score por Modelo e Número de Tópicos}
  \label{fig:figzzz}
\end{figure}

Para esta questão, efetuou-se a escolha do modelo de embeddings mais adequado por meio de validação interna. A Figura~\ref{fig:figzzz} compara seis modelos, em que o eixo horizontal indica o número de tópicos-alvo (\texttt{nr\_topics}) e o eixo vertical apresenta (\(WS\)). Observou-se que o BERTimbau-large supera consistentemente as demais alternativas em todas as faixas, refletindo sua capacidade de gerar vetores semânticos que equilibram coerência interna e diversidade temática. Em contraste, o Legal-BERT registrou desempenho inferior, sugerindo que seu treinamento no domínio jurídico não se generaliza à variedade temática das propostas da plataforma.

Com o BERTimbau-large selecionado como o melhor modelo de embeddings para representar nosso corpus, realizou-se a próxima etapa que buscou a melhor configuração de hiperparâmetros do BERTopic. A Tabela~\ref{tab:experimentos} resume as dez melhores configurações obtidas por meio da busca em grade. Cada configuração apresenta os valores dos seguintes parâmetros: intervalo de n-gramas utilizado no CountVectorizer (\texttt{n\_gram\_range}), número-alvo de tópicos (\texttt{nr\_topics}), tamanho mínimo de tópico (\texttt{min\_topic\_size}), número efetivo de tópicos gerados (\texttt{topics}) e a Pontuação Ponderada Final (\(WS\)).

\begin{table}[ht]
  \centering
  \tiny
  \caption{Top-10 Resultados da Validação Externa}
  \label{tab:experimentos}
  \resizebox{0.90\linewidth}{!}{%
    \begin{tabular}{lrrrccccr}
      \toprule
      \(\texttt{n\_gram\_range}\) & \(\texttt{nr\_topics}\) & \(\texttt{min\_topic\_size}\) & \(\texttt{topics}\) & \(NC\) & \(ND\) & \(WS\) \\
      \midrule
      (1,1) & 70   & 10 & 56.1  & 0,11711 & 0,86234 & 0,26615 \\
      (1,2) & 70   &  3 & 69.0  & 0,08720 & 0,89522 & 0,24880 \\
      (1,1) & 110  & 10 & 72.0  & 0,09156 & 0,85527 & 0,24429 \\
      (1,2) & 90   &  3 & 89.0  & 0,08028 & 0,88826 & 0,24188 \\
      (1,2) & 70   &  5 & 69.0  & 0,08194 & 0,84971 & 0,23549 \\
      (1,2) & 70   & 10 & 69.0  & 0,07758 & 0,84638 & 0,23133 \\
      (1,2) & 110  &  3 & 109.0 & 0,06627 & 0,87550 & 0,22811 \\
      (1,1) & 90   & 10 & 85.4  & 0,07566 & 0,83252 & 0,22702 \\
      (1,1) & 70   &  5 & 69.0  & 0,07143 & 0,83580 & 0,22430 \\
      (1,1) & 130  & 10 & 88.8  & 0,07273 & 0,82973 & 0,22413 \\
      \bottomrule
    \end{tabular}%
  }
\end{table}

Pode-se observar, entre todas as configurações testadas, o melhor resultado foi alcançado quando definimos \texttt{n\_gram\_range = (1,1)}, \texttt{nr\_topics = 70} e \texttt{min\_topic\_size = 10}, produzindo 56 tópicos finais com \(WS = 0{,}2661\). Essa configuração equilibra a necessidade de granularidade ao extrair um número de temas compatível com a diversidade do corpus e a robustez dos tópicos, evitando clusters muito pequenos ou excessivamente dispersos.

\subsection{RQ2 – Em que grau a incorporação de tópicos-semente do VCGE fortalece o alinhamento semântico com as categorias oficiais?}

Para avaliar o quanto a incorporação de tópicos‑semente, extraídos do VCGE, reforça o alinhamento semântico na geração de tópicos com o BERTopic, foram comparados dois cenários: um modelo não supervisionado, sem qualquer reforço de termos, e outra configuração semi‑supervisionada, que incorpora listas de termos‑semente alinhadas às categorias oficiais do VCGE.

A Tabela~\ref{tab:rq3-comparacao} consolida os resultados obtidos em cada cenário para as métricas internas de qualidade de tópicos (NC, ND e WS) e para as métricas externas de alinhamento (ARI e NMI nos níveis hierárquicos N1 e N2), indicando também as diferenças absolutas e percentuais decorrentes da aplicação da semi-supervisão.

\begin{table}[ht]
\centering
\tiny
\caption{Valores de Métricas Internas e Externas para os cenários não supervisionado e semi-supervisionada. $\Delta$ (\%) refere-se à variação percentual de semi-supervisionada em relação a não supervisionado.}
\label{tab:rq3-comparacao}
\resizebox{0.70\linewidth}{!}{
\begin{tabular}{lrrrrr}
\toprule
\textbf{Métrica} & \textbf{Unsup} & \textbf{Semi-sup} & \textbf{Dif.} & \textbf{$\Delta$ (\%)} \\
\midrule
NC & 0,0953         & 0,1166 & +0,0213 & +22,4\% \\
ND & 0,8522 & 0,8420 & \(-\)0,0101 & \(-\)1,2\% \\
WS & 0,2467 & 0,2617 & +0,0150 & +6,1\% \\
ARI (N1) & 0,2095 & 0,3089 & +0,0994 & +47,5\% \\
NMI (N1) & 0,5366 & 0,5495 & +0,0129 & +2,4\% \\
ARI (N2) & 0,2105 & 0,2992 & +0,0887 & +42,1\% \\
NMI (N2) & 0,6088 & 0,6220 & +0,0132 & +2,2\% \\
\bottomrule
\end{tabular}
}
\end{table}

De forma geral, observa‑se que a semi‑supervisão eleva a coerência (NC), enquanto mantém a diversidade (ND) em patamar próximo ao não supervisionado, culminando em um ganho consolidado no WS. No aspecto de alinhamento externo, todos os indicadores (ARI e NMI) apresentam melhorias, especialmente no ARI, que reflete maior aderência às categorias oficiais. 

\begin{figure}[ht]
  \centering
  \includegraphics[width=0.9\linewidth]{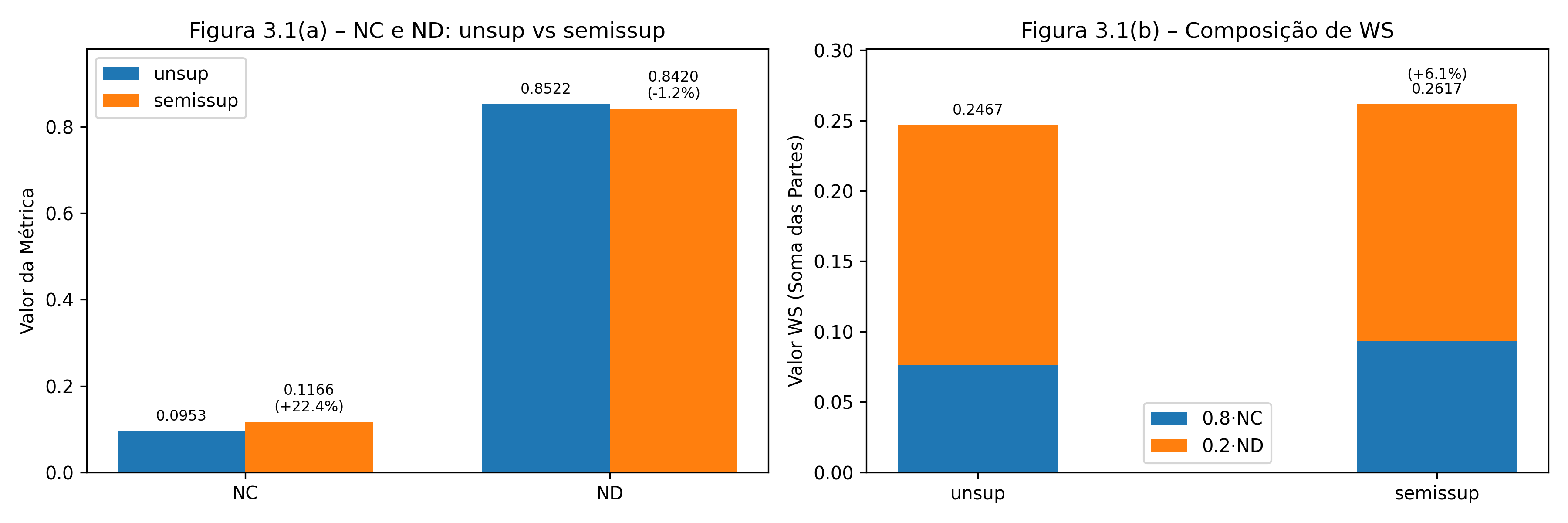}
  \caption{Comparação entre não-supervisionado e semi-supervisionado.}
  \label{fig:rq3-metrica-interna}
\end{figure}

A Figura~\ref{fig:rq3-metrica-interna} nos ajuda a comparar melhor esses resultados obtidos. No lado esquerdo, as barras ilustram a diferença entre NC e ND ao adotar tópicos-semente. Já o lado direito detalha a composição de \(WS\) em cada cenário, considerando a ponderação entre as métricas internas também para cada cenário executado.

Pode-se observar que a medida de NC passa de 0.0953 no cenário não supervisionado para 0.1166 quando aplicamos a semi‑supervisão com tópicos-semente, o que equivale a um aumento de 22.4\%. Já a métrica ND, por sua vez, sofre uma leve queda de 0.8522 para 0.8420 ($-1{.}2$\%). Quando combinamos essas duas medidas através do \(WS\), observamos um crescimento de 0.2467 para 0.2617, ou seja, +6,1\%, indicando que o ganho em coerência supera a pequena perda em diversidade.

\begin{figure}[ht]
  \centering
  \includegraphics[width=0.9\linewidth]{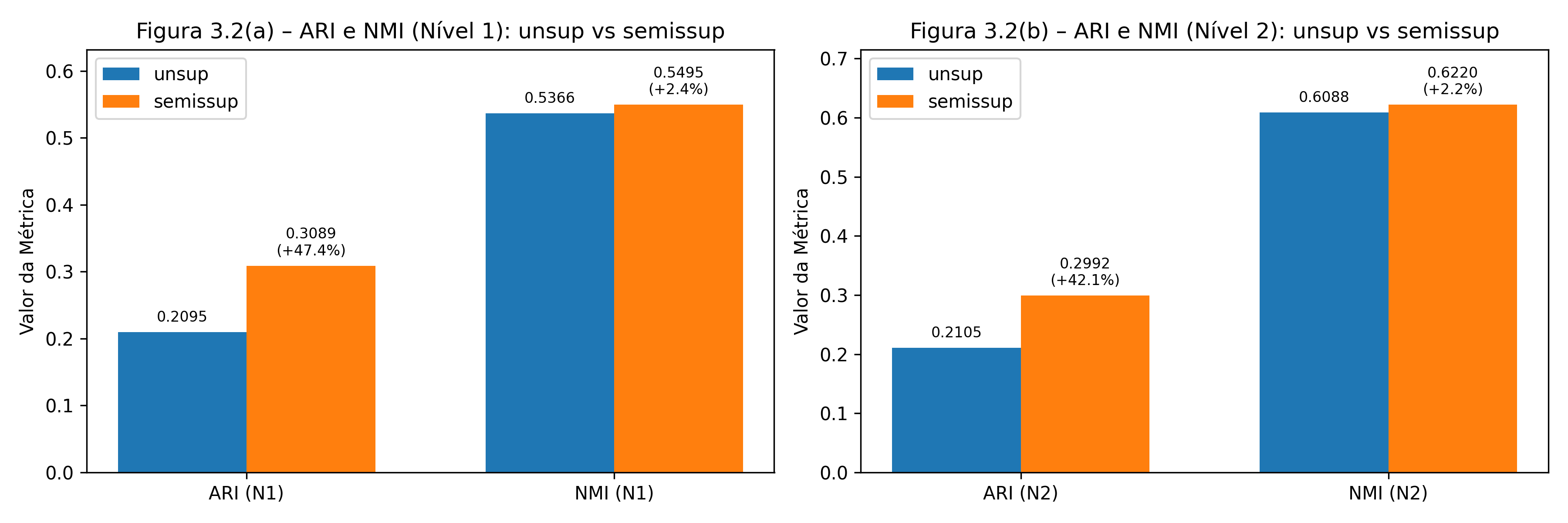}
  \caption{Métricas externas de alinhamento (ARI e NMI) nos níveis N1 e N2.}
  \label{fig:rq3-metrica-externa}
\end{figure}

A Figura~\ref{fig:rq3-metrica-externa} ilustra o efeito da inclusão de tópicos‑semente nas métricas de alinhamento dos clusters gerados. À esquerda, são comparadas as correspondências entre tópicos e categorias de nível N1 do VCGE; à direita, o mesmo exercício para as subcategorias de nível N2.

Quanto às métricas externas, que quantificam o alinhamento entre tópicos gerados e categorias oficiais do VCGE, o ARI no nível N1 salta de 0.2095 para 0.3089 (+47.5\%), enquanto o ARI no nível N2 cresce de 0.2105 para 0.2992 (+42.1\%). Esses aumentos demonstram que a semi‑supervisão aproxima substancialmente os clusters das categorias gerais (N1) e das subcategorias (N2) definidas pelo VCGE. A NMI no nível N1 passa de 0,5366 para 0,5495 (+2,4\%) e a NMI no nível N2 de 0,6088 para 0,6220 (+2,2\%), refletindo um alinhamento mais consistente em termos de informação mútua normalizada.

Portanto, a semi-supervisão por \emph{tópicos-semente} fortalece de modo mensurável o mapeamento semântico entre os clusters gerados e as categorias oficiais do VCGE, equilibrando coerência interna e diversidade temática sem comprometer a cobertura da taxonomia institucional.

\section{Discussões}
\subsection{Estratégia e Aplicabilidade}

Os resultados indicam que a combinação de embeddings especializados em português brasileiro com semi-supervisão baseada em vocabulários oficiais é a estratégia mais eficaz para transformar grandes volumes de propostas cidadãs em insumos acionáveis: a adoção do BERTimbau-large para gerar representações semânticas, aliada ao uso de \emph{palavras-semente} do VCGE, produziu tópicos que conciliam riqueza linguística e alinhamento institucional. Em termos operacionais, esse pipeline pode ser integrado imediatamente a plataformas de engajamento digital. Ao automatizar a categorização, ele reduz substancialmente o trabalho manual, libera especialistas para análises aprofundadas e permite a detecção em tempo real de tendências emergentes (por exemplo, picos de interesse em saúde ou meio ambiente).

\subsection{Impactos e Beneficiários}

Este método oferece dois benefícios principais: acelera a classificação de propostas, otimizando tempo e recursos humanos; e eleva a qualidade das informações, pois as categorias geradas são consistentes com taxonomias oficiais, garantindo rastreabilidade. Os principais beneficiários são equipes de formulação de políticas públicas, que passam a dispor de relatórios temáticos confiáveis para embasar decisões estratégicas, e desenvolvedores de plataformas cívicas, que podem integrar um módulo plugável de navegação temática para aprimorar interfaces de consulta e engajamento. Além disso, o alinhamento com vocabulários institucionais reforça a legitimidade do processo participativo: quando cidadãos percebem que suas contribuições são corretamente reconhecidas e agrupadas em categorias oficiais, fortalece-se a confiança e o engajamento democrático.

\subsection{Replicabilidade e Evolução Contínua}

Este estudo demonstra a viabilidade de IA orientada por vocabulários governamentais: mais do que extrair padrões estatísticos, o pipeline respeita e reforça estruturas institucionais. A metodologia proposta é replicável em diferentes contextos estaduais, municipais ou setoriais sem grandes ajustes, graças ao uso de embeddings em português e vocabulários oficiais amplamente disponíveis. Por fim, introduz-se a perspectiva de evolução contínua por meio de ciclos de feedback humano-modelo. Avaliações periódicas de especialistas podem refinar automaticamente as seed words e ajustar parâmetros, mantendo o sistema alinhado às mudanças nas demandas cidadãs. Essa abordagem garante adaptabilidade e relevância em um ambiente de participação digital dinâmico.

\section{Trabalhos Relacionados}

A aplicação de técnicas de Processamento de Linguagem Natural em documentos  administrativos tem ganhado destaque, com evidências de que modelos especializados aumentam a performance em diversas tarefas. \cite{silveira2021topic} demonstraram que o uso do LEGAL-BERT em decisões judiciais melhora a coerência dos tópicos extraídos. \cite{silva2022lipset} desenvolveram o LiPSet, corpus anotado de licitações, enquanto \cite{constantino2022segmentaccao} aplicaram aprendizado ativo na segmentação de diários oficiais, alcançando 85\% de acurácia com menos dados rotulados. No pré-treinamento adaptativo de domínio (DAPT), \cite{silva2024evaluating} mostraram que corpora alinhados ao domínio governamental elevam a precisão de modelos baseados em BERT, reforçando a importância da escolha do conjunto de dados. Complementarmente, \cite{hott2023evaluating} compararam embeddings como BERTimbau, LaBSE e LiBERT-SE em tópicos de compras públicas, evidenciando a vantagem de modelos em português. Seguindo essa linha, \cite{silva2024govbert} apresentaram o GovBERT-BR, treinado com textos oficiais de órgãos públicos brasileiros, que superou modelos generalistas em tarefas de classificação e segmentação. Apesar desses avanços, ainda há pouca investigação sobre modelagem de tópicos em plataformas participativas de larga escala e sua relação com taxonomias institucionais; lacuna que este trabalho busca preencher por meio de um pipeline com validação cruzada e uso de vocabulários oficiais.

\section{Conclusão}

Este trabalho propôs e avaliou um pipeline de modelagem de tópicos para a classificação de propostas públicas submetidas à plataforma Brasil Participativo. Ao combinar BERTopic com quatro diferentes embeddings, ajustes de hiperparâmetros e conhecimento institucional incorporado por meio de \emph{palavras-semente} extraídos do VCGE, foi possível elevar a coerência e o alinhamento semântico dos tópicos gerados sem comprometer sua diversidade. A utilização de LLMs para rotulação automática mostrou-se eficaz para reduzir o esforço manual de validação, favorecendo a escalabilidade e a eficiência do processo. Ainda assim, mantém-se a necessidade de validação humana para casos de baixa confiança e a adaptação do vocabulário-semente e dos parâmetros quando o método for transferido para outros contextos. Em suma, o pipeline oferece uma solução operacional para transformar contribuições cidadãs em insumos acionáveis para a formulação de políticas públicas.

\bibliographystyle{sbc}
\bibliography{sbc-template}

\end{document}